\documentclass{ws-mpla}
\usepackage[super]{cite}
\usepackage{graphicx}

\begin{document}

\markboth{Yevgeny.~V.~Stadnik \and Victor.~V.~Flambaum}
{New generation low-energy probes for ultralight axion and scalar dark matter}


\title{New generation low-energy probes for ultralight axion and scalar dark matter}

\author{Yevgeny V.~Stadnik$^{1}$ \and Victor V.~Flambaum$^{1,2}$}

\address{$^{1}$School of Physics, University of New South Wales, Sydney 2052, Australia}
\address{$^{2}$Mainz Institute for Theoretical Physics, Johannes Gutenberg University Mainz, D 55122 Mainz, Germany}

\maketitle


\begin{abstract}
We present a brief overview of a new generation of high-precision laboratory and astrophysical measurements to search for ultralight (sub-eV) axion, axion-like pseudoscalar and scalar dark matter, which form either a coherent condensate or topological defects (solitons). In these new detection methods, the sought effects are linear in the interaction constant between dark matter and ordinary matter, which is in stark contrast to traditional searches for dark matter, where the sought effects are quadratic or higher order in the underlying interaction constants (which are extremely small).

\end{abstract}

\ccode{PACS Nos.: 95.35.+d, 14.80.Va, 11.27.+d, 07.55.Ge, 06.30.Ft, 11.30.Er, 06.20.Jr}

\emph{\textbf{Introduction}} ---
Dark matter (DM) remains one of the most important unsolved problems in contemporary physics. DM is a non-luminous, non-baryonic form of matter than interacts very weakly with itself and Standard Model (SM) matter. Evidence for the existence of DM includes the observed flatness of galactic rotation curves to large distances away from the galactic centre \cite{Zwicky1933,Rubin1970}, gravitational lensing observations of the Bullet Cluster \cite{Bullet2003}, angular fluctuations in the cosmic microwave background (CMB) spectrum \cite{CMB2009} and the need for non-baryonic matter to explain observed structure formation \cite{Silk2005Review}. Observations of stellar orbital velocities in our local galactic neighbourhood give the cold DM energy density within our local galactic neighbourhood of \cite{PDG2014}:
\begin{equation}
\label{rho_CDM_local}
\rho_{\textrm{CDM}}^{\textrm{local}} \simeq 0.4~\textrm{GeV/cm}^3 ,
\end{equation}
while the latest Wilkinson Microwave Anisotropy Probe (WMAP) observations give a present-day mean DM energy density of \cite{PDG2014}:
\begin{equation}
\label{rho_DM_mean-WMAP}
\bar{\rho}_{\textrm{DM}} = 1.3 \times 10^{-6}~\textrm{GeV/cm}^3 .
\end{equation}
Despite the overwhelming evidence for its existence, direct searches for DM have not yet produced a strong positive result, leaving the identity and non-gravitational properties of DM in a state of mystery. Traditional searches for the scattering of weakly interacting massive particle (WIMP) DM particles off nuclei (see e.g.~Refs.~\cite{SuperCDMS2014,CoGeNT2011,CRESST2014,DAMA2008,LUX2013,XENON2013}) are all based on effects that are fourth order in the underlying interaction parameters, which are extremely small. In the present review, we present a brief overview of a new generation of methods to search for ultralight axion and scalar DM that are based on effects that are \emph{\textbf{first order}} in the interaction strength between these DM candidates and SM particles. 
\emph{\textbf{The strong CP problem and the QCD axion}} --- 
When the SM was been developed during the 1970s, it became apparent that there was an issue in the Quantum Chromodynamics (QCD) sector as far the combined charge-parity ($CP$) symmetry was concerned. The QCD Lagrangian contains the $P$,$CP$-violating term \cite{Polyakov1975,Hooft1976,Jackiw1976,Gross1976}: 
\begin{equation}
\label{theta-term}
\mathcal{L}_{\textrm{QCD}}^{\theta} = \theta \frac{g^2}{32 \pi^2} G\tilde{G} ,
\end{equation}
where $\theta$ is the angle that quantifies the amount of $CP$ violation within the QCD sector, $g^2/4\pi = 14.5$ is the colour coupling constant, and $G$ and $\tilde{G}$ are the gluonic field tensor and its dual, respectively. Account of weak interaction effects results in a shift of $\theta$ from its bare value to the observable value $\bar{\theta}$ \cite{Peccei1981}. The angle $\bar{\theta}$ may in principle have assumed any value in the range $-\pi \le \bar{\theta} \le +\pi$, but its observed value from measurements of the permanent static neutron electric dipole moment (EDM) is constrained to be $|\bar{\theta}| < 10^{-10}$ \cite{Baker2006NEDM}. The smallness of the observed value of $\bar{\theta}$ constitutes the strong $CP$ problem. An elegant and the most widely accepted resolution of the strong $CP$ problem was proposed by Roberto Peccei and Helen Quinn \cite{PQ1977A,PQ1977B}, in which the $\theta$ parameter was interpretted as a dynamical field (the massive pseudoscalar axion, $a$): $\bar{\theta} \to a(t)/f_a$, where $f_a$ is the axion decay constant. Initially, the axion field is constant ($\bar{\theta} \sim 1$ at times when $m_a \ll H(t)$), but for times when $m_a \gg H(t)$, where $H(t)$ is the Hubble constant, the axion undergoes oscillations about the minimum of its potential, which corresponds to $\bar{\theta}=0$, hence alleviating the strong $CP$ problem \cite{Wilczek1983,Abbott1983,Dine1983}.

Although the original PQWW model of the axion \cite{PQ1977A,PQ1977B,Weinberg1978,Wilczek1978} was quickly ruled out experimentally, the KSVZ \cite{Kim1979,Shifman1980} and DFSZ \cite{Zhitnitsky1980,Dine1981} axion models turned out to be compatible with all terrrestrial and astrophysical observations (for some of the more recent invisible axion models based on the Peccei-Quinn symmetry, we refer the reader to Refs.~\cite{Lindner2010PQ,Celis2014PQ,Celis2015PQ,Anh2015PQ}). The properties of the QCD axion are predominantly determined by the axion decay constant $f_a$. In particular, the QCD axion mass $m_a$ is related to $f_a$ via the relation
\begin{equation}
\label{QCD_axion_mass}
m_a \simeq 6 \times 10^{-5}~\textrm{eV} \left( \frac{10^{11}~\textrm{GeV}}{f_a} \right) .
\end{equation}


\emph{\textbf{Axions as cold dark matter}} --- 
It has widely been noted that the QCD axion, as well as axion-like pseudoscalar particles (ALPs), for which no predictive mass-coupling constant relation akin to Eq.~(\ref{QCD_axion_mass}) is guaranteed, may be promising cold DM candidates. Astrophysical and cosmological constraints on axion parameters \cite{Khlopov1978,Wilczek1983,Abbott1983,Dine1983,Raffelt2008,Marsh2013DMHalo,Wong2013,Blum2014BBN,Marsh2015ULA_reionise,Marsh2015ULA_cosmo} greatly assist in laboratory searches for axions. Traditional haloscope (ADMX) \cite{ADMX2010} and helioscope (CAST) \cite{CAST2014} experiments that search for galactic and solar axions \cite{Sikivie1983ADMX}, respectively, have shed valuable light on our understanding of the possible axion parameter space for the axion-photon coupling. The next-generation IAXO helioscope experiment \cite{IAXO2014} will continue the search for solar axions. Searches for solar axions via the axio-electric effect with scintillator detectors have also been conducted \cite{Avignone1987,Dzuba2010AE,Derbin2012}. Various `light-shining-through-wall' \cite{GammeV2008,ALPS2013,CROWS2013} and vacuum birefringence \cite{BMV2007,PVLAS2013} searches for axions and ALPs via the axion-photon coupling have also been performed. 

Searches for ultralight (sub-eV) axions, ALPs and scalars in tabletop experiments via the macroscopic forces they produce due to their couplings with the electron and nucleons were first proposed in Ref.~\cite{Wilczek1983_New-force}. Atomic magnetometry \cite{Lamoreaux1996_NF,Walsworth2008_NF,Romalis2009_NF,Petukhov2010_NF,Tullney2013_NF,Chu2013_NF,Bulatowicz2013_NF,Kimball2015_NF}, torsion pendulum \cite{Heckel2011_NF}, differential force measurements \cite{Mostepanenko2015_NF} and ultracold neutron experiments \cite{Baessler2007_NF,Serebrov2010_NF,Grenoble2015_NF} have collectively probed the axion-electron and axion-nucleon couplings over an expansive range of axion masses (see also Ref.~\cite{Raffelt2012} for constraints on axion-electron and axion-nucleon interactions from a combination of terrestrial equivalence principle tests and astrophysical stellar energy-loss bounds). A number of new proposals to search for axions via effects that, like traditional searches, are quadratic or higher order in the combination of axion parameters $a/f_a$ ($a/f_a \simeq 4 \times 10^{-19}$ for the QCD axion, which obeys relation (\ref{QCD_axion_mass}) and which also saturates the local cold DM energy density in Eq.~(\ref{rho_CDM_local})) have been put forward over the recent years, see e.g.~Refs.~\cite{Baker2012,Rybka2015,Ringwald2013,Beck2013,Beck2015,Popov2014,Arvanitaki2014NMR,Sikivie2014Atomic,Arias2015}.

\emph{\textbf{Ultralight axion and scalar dark matter}} --- 
The number density of ultralight (sub-eV) bosonic DM particles per de Broglie volume can greatly exceed unity, $n/\lambda_{\textrm{dB}}^3 \gg 1$. As a result, ultralight scalar (and axion) DM readily forms a coherently oscillating condensate
\begin{equation}
\phi(\mathbf{r},t) = \phi_0 \cos( \varepsilon_\phi t - \mathbf{p}_\phi \cdot \mathbf{r} ) ,
\end{equation}
with an amplitude $\phi_0 \simeq \sqrt{2 \rho_{\textrm{scalar}}}/ m_\phi$, where $\rho_{\textrm{scalar}}$ is the energy density associated with the scalar DM field, and with an oscillation frequency set by the mass of the DM particle (in the natural units $\hbar = c =1$): $\omega_\phi \simeq m_\phi$, for non-relativistic cold DM.
The allowed mass of bosonic DM particles covers a very wide range. For elementary bosonic DM, the upper mass limit is set by the requirement that these particles do not form elementary black holes: $m_\phi < 10^{28}~\textrm{eV}$ (although this upper limit is relaxed if DM does not consist of elementary particles). The simplest lower limit arises from the requirement that the de Broglie wavelength of the DM particles not exceed the halo size of the smallest galaxies, giving $m_\phi \gtrsim 10^{-22}$ eV. This simple estimate is in fact in good agreement with more rigorous limits obtained from cosmological and astrophysical investigations, based on Lyman-alpha forest observations \cite{Marsh2013DMHalo,Viel2013warmDM}, observed high-redshift galaxy formation \cite{Marsh2015ULA_reionise} and CMB observations \cite{Marsh2015ULA_cosmo}. Due to its effects on structure formation, ultralight scalar and axion DM in the mass range $10^{-24} - 10^{-20}$ eV has been proposed \cite{Gruzinov2000fuzzyDM,Marsh2013DMHalo} to solve several long-standing astrophysical puzzles, such as the cusp-core, missing satellite and too-big-to-fail problems \cite{Weinberg2015ReviewDMIssues} (see also the earlier work of Ref.~\cite{Khlopov1985scalar}). 

Another possibility for ultralight bosonic DM is the formation of topological defects (solitons), which arise from the stabilisation of a DM field under a suitable self-potential \cite{tHooft1974,Polyakov1974,Kibble1976,Khlopov1978B,Zeldovich1980,Vilenkin1982,Sikivie1982}. Topological defects, which make up a sub-dominant fraction of DM, may function as seeds for structure formation \cite{Brandenberger1987}. For some of the more recent developments on topological defects, we refer the reader to Refs.~\cite{Kirk2014,Sugiyama2015,Harko2015,Marsh2015,Oksanen2015}~, while for the classical review, we refer the reader to Ref.~\cite{Vilenkin1985REVIEW}.

\emph{\textbf{{Axion dark matter searches}}} --- 
The axion couples to SM particles as follows (we consider only couplings that are of direct interest to experimental searches):
\begin{align}
\label{Axion_couplings}
\mathcal{L}_{\textrm{axion}} = \frac{a}{f_a} \frac{g^2}{32\pi^2} G\tilde{G} + \frac{C_\gamma a}{f_a} \frac{e^2}{32\pi^2} F\tilde{F} - \sum_{f} \frac{C_f}{2f_a} \partial_\mu a ~ \bar{f} \gamma^\mu \gamma^5 f ,
\end{align}
where the first term represents the coupling of the axion field to the gluonic field tensor $G$ and its dual $\tilde{G}$, the second term represents the coupling of the axion field to the electromagnetic field tensor $F$ and its dual $\tilde{F}$, while the third term represents the coupling of the derivative of the axion field to the fermion axial-vector currents $\bar{f} \gamma^\mu \gamma^5 f$. $C_\gamma$ and $C_f$ are model-dependent coefficients. Typically, $|C_\gamma| \sim 1$ and $|C_n| \sim |C_p| \sim 1$ in models of the QCD axion \cite{Kim2010RMP-axions,Srednicki1985axion}. Within the DFSZ model, where the tree level coupling of the axion to the electron is non-vanishing, $|C_e| \sim 1$ \cite{Srednicki1985axion}. However, within the KSVZ model, $|C_e| \sim 10^{-3}$, since the tree level coupling vanishes and the dominant effect arises at the 1-loop level \cite{Srednicki1985axion}. For ALPs, the coefficients $C_\gamma$ and $C_f$ are essentially free parameters, and the coupling to gluons is presumed absent.

\emph{`Axion wind' effect} --- The spatial terms in the coupling of the derivative of an axion condensate to the fermion axial-vector currents in Eq.~(\ref{Axion_couplings}) lead to the following non-relativistic Hamiltonian \cite{Flambaum2013Patras,Stadnik2014axions,Graham2013}
\begin{equation}
\label{axion_wind}
H_{\textrm{eff}}(t) = \frac{C_f a_0}{2f_a} \sin(m_a t) ~ p_a \cdot \sigma_{f} ,  
\end{equation}
which implies that a spin-polarised source of particles interacts with the axion 3-momentum, producing oscillating shifts in the energy of the spin-polarised source at two characteristic frequencies: $\omega_1 \simeq m_a$ and $\omega_2 = 2\pi/T_{\textrm{sidereal}}$, where $T_{\textrm{sidereal}} = 23.93$ hours is the sidereal day duration. This is the `axion wind' effect, which may be sought for using a variety of spin-polarised sources, for example, atomic co-magnetometers, torsion pendula and ultracold neutrons. Distortion of the axion condensate by the gravitational field of a massive body, such as the Sun or Earth, results in an additional axion-induced oscillating spin-gravity coupling \cite{Stadnik2014axions}. For couplings of the axion to nucleons inside the nucleus, isotopic dependence ($C_n \ne C_p$) requires knowledge of the proton and neutron spin contributions. The proton and neutron spin contributions for nuclei of experimental interest have been calculated in Ref.~\cite{Stadnik2015NMBE}.

\emph{Transient `axion wind' effect} --- While stable domain wall structures that consist of the QCD axion would lead to disastrous consequences in cosmology by storing too much energy \cite{Sikivie1982}, domain walls and other topological defect structures consisting of scalars or ALPs are viable for certain combinations of parameters. Topological defects that consist of ALPs may interact with fermion axial-vector currents via the derivative coupling in Eq.~(\ref{Axion_couplings}), which in the non-relativistic limit reads \cite{GNOME2013A}
\begin{equation}
\label{axion_wind_transient}
H_{\textrm{eff}}(t) = \frac{C_f}{2f_a} \left( \mathbf{\nabla} a \right) \cdot \sigma_{f} .
\end{equation}
Eq.~(\ref{axion_wind_transient}) implies that a spin-polarised source of particles will temporarily interact with a topological defect as the defect passes through the system.  A global network of detectors, such as atomic co-magnetometers, \cite{GNOME2013A,GNOME2013B} has been proposed to detect these correlated transient-in-time signals, produced by the passage of a defect through Earth.

\emph{Oscillating $P$,$T$-odd electromagnetic moments} --- Interaction of a QCD axion condensate with the gluon fields in Eq.~(\ref{Axion_couplings}) produces an oscillating neutron electric dipole moment \cite{Graham2011}
\begin{equation}
\label{Osc_NEDM}
d_n = 1.2 \times 10^{-16} \frac{a_0 \cos(m_a t)}{f_a} ~ e~\textrm{cm} ,
\end{equation}
which induces oscillating nuclear Schiff moments \cite{Graham2011,Stadnik2014axions} and oscillating nuclear magnetic quadrupole moments \cite{Roberts2014long}. In nuclei, the dominant mechanism for the induction of oscillating $P$,$T$-odd electromagnetic moments by axions comes from the $P$,$T$-violating nucleon-nucleon interaction that is mediated by pion exchange (with the axion condensate supplying the oscillating source of $P$ and $T$ violation at one of the $\pi N N$ vertices) \cite{Stadnik2014axions,Roberts2014long}. 
Oscillating nuclear Schiff moments are observable with diamagnetic species, while nuclear magnetic quadrupole moments require paramagnetic species. Resonance magnetometry searches in diamagnetic solid-state media have proposed to search for axion-induced oscillating nuclear Schiff moments \cite{CASPER2014}.

The temporal term in the coupling of the derivative of an axion condensate to the axial-vector currents of atomic or molecular electrons in Eq.~(\ref{Axion_couplings}) produces oscillating electric dipole moments in paramagnetic species \cite{Stadnik2014axions,Roberts2014long,Roberts2014prl}. In the non-relativistic approximation, the axion-induced oscillating electric dipole moment of a paramagnetic atom with a single valence electron in the $s$-wave is \cite{Stadnik2014axions}
\begin{equation}
\label{paramagnetic_EDM}
d_a = - \frac{3 C_e a_0 m_a^2 \alpha_s}{2 f_a \alpha} e \cos(m_a t) ,
\end{equation}
where $\alpha_s$ is static atomic scalar polarisability and $\alpha \simeq 1/137$ is the electromagnetic fine-structure constant. Fully relativistic calculations have been performed in Refs.~\cite{Roberts2014long,Roberts2014prl}, in which the relativistic corrections to Eq.~(\ref{paramagnetic_EDM}) were found to be $\sim 10\%$ for moderately heavy atoms ($Z \sim 50$).

\emph{Oscillating electric current flows along magnetic field lines} --- In the presence of an externally applied magnetic field $\mathbf{B}_0$, an axion condensate that interacts with the electromagnetic field in Eq.~(\ref{Axion_couplings}) induces an electric current density 
\begin{equation}
\label{osc_current}
\mathbf{j}_a = \frac{\alpha C_\gamma a_0  m_a}{\pi f_a} \mathbf{B}_0 \sin(m_a t) ,
\end{equation}
which in turn produces a magnetic field $\mathbf{B}_a$ that satisfies $\nabla \times \mathbf{B}_a = \mathbf{j}_a$, and which can be amplified with an LC circuit and subsequently detected using a magnetometer \cite{Sikivie2014LC}. An analogous strategy has also been proposed for the detection of hidden photons \cite{Dobrich2014B}.

\emph{\textbf{Scalar dark matter searches}} ---
Scalar DM may interact linearly with SM particles as follows
\begin{align}
\label{Scalar_couplings_lin}
\mathcal{L}_{\textrm{scalar}}^{\textrm{lin.}} = \frac{\phi}{\Lambda_\gamma} \frac{F_{\mu \nu} F^{\mu \nu}}{4} - \sum_{f} \frac{\phi}{\Lambda_f} m_f \bar{f} f + \sum_{V} \frac{\phi}{\Lambda_V} \frac{M_V^2}{2} V_\nu V^\nu ,
\end{align}
where the first term represents the coupling of the scalar field to the electromagnetic field tensor $F$, the second term represents the coupling of the scalar field to the fermion bilinears $\bar{f} f$, while the third term represents the coupling of the scalar field to the massive vector boson wavefunctions. The new physics energy scales $\Lambda_X$ that appear in (\ref{Scalar_couplings_lin}) 
are known to be very large energy scales from equivalence principle tests, which include lunar laser ranging \cite{Turyshev2004,Turyshev2012} and the E\"{o}tWash experiment \cite{Adelberger2008,Adelberger2012} (see also Ref.~\cite{Raffelt1999} for constraints from stellar energy loss bounds). The couplings in Eq.~(\ref{Scalar_couplings_lin}) alter the electromagnetic fine-structure constant $\alpha$ and particle masses as follows
\begin{align}
\label{Lin_constants}
\alpha \to \frac{\alpha}{1-\phi/\Lambda_\gamma} \simeq \alpha \left[1 +\frac{\phi}{\Lambda_\gamma} \right] , \frac{\delta m_f}{m_f} = \frac{\phi}{\Lambda_f} , \frac{\delta M_V}{M_V} = \frac{\phi}{\Lambda_V} .
\end{align}
Likewise, scalar DM may also interact quadratically with SM particles as follows
\begin{align}
\label{Scalar_couplings_quad}
\mathcal{L}_{\textrm{scalar}}^{\textrm{quad.}} = \frac{\phi^2}{(\Lambda'_\gamma)^2} \frac{F_{\mu \nu} F^{\mu \nu}}{4} - \sum_{f} \frac{\phi^2}{(\Lambda'_f)^2} m_f \bar{f} f + \sum_{V} \frac{\phi^2}{(\Lambda'_V)^2} \frac{M_V^2}{2} V_\nu V^\nu ,
\end{align}
where the new physics energy scales $\Lambda'_X$ are much less severely constrained than the corresponding $\Lambda_X$, from astrophysical observations, most notably bounds from supernova energy loss, and equivalence principle tests \cite{Olive2008}. The couplings in Eq.~(\ref{Scalar_couplings_quad}) alter $\alpha$ and the particle masses as follows
\begin{align}
\label{Quad_constants}
\alpha \to \frac{\alpha}{1-\phi^2/(\Lambda'_\gamma)^2} \simeq \alpha \left[1 +\frac{\phi^2}{(\Lambda'_\gamma)^2} \right] , \frac{\delta m_f}{m_f} = \frac{\phi^2}{(\Lambda'_f)^2} , \frac{\delta M_V}{M_V} = \frac{\phi^2}{(\Lambda'_V)^2} .
\end{align}

\emph{Oscillating variations of the fundamental constants} --- The interactions of a scalar condensate with the SM fields via the linear couplings in Eq.~(\ref{Scalar_couplings_lin}) \cite{Arvanitaki2015scalar,Stadnik2015laser,Stadnik2015BBN} and via the quadratic couplings in Eq.~(\ref{Scalar_couplings_quad}) \cite{Stadnik2015laser,Stadnik2015BBN,Stadnik2015DM-VFCs} produce oscillating variations of the fundamental constants and particle masses, which can be sought for with high-precision terrestrial experiments involving atomic clocks \cite{Arvanitaki2015scalar,Stadnik2015BBN,Stadnik2015DM-VFCs} and laser interferometers \cite{Stadnik2015laser}. A multitude of atomic, highly-charged-ionic, molecular and nuclear systems can be used in clock-based searches, see the reviews \cite{Flambaum2008clock_review,Flambaum2014HCI_review} for summaries of the possible systems. The first laboratory clock-based search for oscillating variation of $\alpha$ has very recently been completedly \cite{Budker2015scalar}, and the results of this search have been used to place stringent constraints on the photon interaction parameters $\Lambda_\gamma$ \cite{Budker2015scalar} and $\Lambda'_\gamma$ \cite{Stadnik2015BBN,Stadnik2015DM-VFCs} for the range of scalar DM masses $10^{-24}~\textrm{eV} \lesssim m_\phi \lesssim 10^{-16}~\textrm{eV}$.

\emph{`Slow' drifts of the fundamental constants} --- The interaction of a scalar condensate with the SM fields via the quadratic couplings in Eq.~(\ref{Scalar_couplings_quad}) produces not only oscillating variations of the fundamental constants and particle masses, but also `slow' drifts of the fundamental constants and particle masses through the terms involving $\left< \phi^2 \right> \ne 0$, which is related to the local ambient energy density of scalar DM \cite{Stadnik2015BBN,Stadnik2015DM-VFCs}. Measurements and SM calculations of the neutron-to-proton ratio at the time of the weak interaction freeze-out prior to Big Bang nucleosynthesis (BBN) give stringent constraints on the photon, quark and massive vector boson interaction parameters: $\Lambda'_\gamma$, $\Lambda'_q$ and $\Lambda'_V$, for the range of scalar DM masses $m_\phi \lesssim 10^{-4}$ eV \cite{Stadnik2015BBN,Stadnik2015DM-VFCs}. CMB measurements constrain the photon and electron interaction parameters: $\Lambda'_\gamma$ and $\Lambda'_e$, for the range of scalar DM masses $m_\phi \lesssim 10^{-4}$ eV \cite{Stadnik2015DM-VFCs}. Measurements and SM calculations of the neutron-to-proton ratio at the time of the weak interaction freeze-out prior to BBN also give constraints on the interaction parameters $\Lambda_\gamma$, $\Lambda_q$ and $\Lambda_V$, for the range of scalar DM masses $m_\phi \lesssim 10^{-16}$ eV, when the scalar condensate had not yet begun to oscillate \cite{Stadnik2015BBN}.

\emph{Transient-in-time variations of the fundamental constants} --- The interactions of a scalar DM field, which comprises a topological defect, with SM fields via either of the couplings in Eqs.~(\ref{Scalar_couplings_lin}) or (\ref{Scalar_couplings_quad}) alter the fundamental constants and particle masses inside the defect, giving rise to local transient-in-time variations as a defect temporarily passes through a region of space \cite{Derevianko2014,Stadnik2014defects}. These transient-in-time variations can be sought for using a global network of detectors, including atomic clocks \cite{Derevianko2014} and laser interferometers \cite{Stadnik2015laser}, as well as a network of pulsars \cite{Stadnik2014defects}, such as the international pulsar timing array \cite{PTA2010}. For non-adiabatic passage of a defect (involving a sufficiently narrow and/or rapidly travelling defect) through a pulsar, a topological defect may trigger a pulsar glitch \cite{Stadnik2014defects}. Numerous pulsar glitches have already been observed (see e.g.~Ref.~\cite{Lyne2006_pulsar_book} for an overview), but their underlying cause is still disputed (see e.g.~Ref.~\cite{Haskell2015_glitch_review} for a recent review).

\emph{\textbf{Outlook}} ---
Effects that are linear in the interaction constant between DM and ordinary matter provide strong motivation for a new generation of searches for ultralight axion and scalar DM. The first such laboratory search for ultralight scalar DM using atomic spectroscopy measurements in dysprosium has recently been completed \cite{Budker2015scalar}. A number of new generation laboratory searches for ultralight axion and scalar DM using atomic and solid-state magnetometry, atomic clocks, interferometry, torsion pendula and ultracold neutrons are either already in progress or planned to commence in the near future.

\emph{\textbf{{Acknowledgments}}} ---
We would like to thank Maxim Yu.~Khlopov for the invitation to write this review. This work was supported by the Australian Research Council.

\end{document}